# Elastic Properties of Carbon nanotubes : An atomistic approach

Roby Cherian and Priya Mahadevan*

S.N. Bose National Centre for Basic Sciences
JD Block, Sector III, Salt Lake,
Kolkata – 700098

Energetically the single sheet of graphite (graphene) is more stable than the nanotube. The energy difference between the two systems can be directly related to the strain energy involved in rolling up the graphene sheet to form the nanotube. We have carried out first-principle electronic structure calculations and evaluated the strain energy as a function of the nanotube radius. The dependence of the strain energy on the diameter of the nanotube has been found by several groups to be well-described by a continuum elasticity model. We attempt to examine why this is the case and show where atomistics enter the description.

**Keywords:** ELASTICITY; NANOTUBES

**Introduction**

The discovery of carbon nanotubes (CNT) was a serendipitous one where these materials were found first as a byproduct during the preparation of bulk quantities of fullerene clusters [1]. Since then these materials have moved a long way and are now considered to be the building blocks of future nanoscale electronic and mechanical devices [2]. Neglecting the structure at the ends, a CNT consists of graphitic sheets which have been rolled into cylindrical shape. This however does not happen under ambient conditions [3] as there is an energy cost involved in bending the perfect $sp^2$ bonds that exist in a single sheet of graphene to form the bent bonds of a seamless cylinder. Usually the presence of a catalyst is required, though the actual role of the catalyst is not well understood.

An important feature of these materials that emerges from the strong $sp^2$ bonding network is that they are stiffer than steel and resistant to damage from physical forces. Pressing the tip of the nanotube will cause it to bend without damage to the tip or the whole CNT. When the force is removed, the tip of the nanotube will recover its original state.

In the first part of our study we have examined the energy cost involved in rolling up a single graphite sheet to form a nanotube. This would indeed depend on the radius of the tube, as the energy cost would primarily stem from the cost involved in bending the C-C bond from its ideal value for a sheet of graphite. This would go as $\cos^2\theta$ where $\theta$ is the C-C-C angle being changed in the process. Indeed our results find a dependence on the angle. The dependence on the angle translates into a $1/R^2$ dependence of the strain energy, where R is the radius of the tube. In contrast to such an approach which approaches the problem from the microscopics, there have been several groups [4] who have approached the problem from the continuum limit, totally ignoring the atomistics. This approach in the case of CNTs gives the same dependence as what is found in a completely atomistic approach. We go on to clarify why this is the case.

We calculate the equilibrium bond length in both graphite and diamond. This is found to be 1.43 Å for graphite. The diamond structure is found to be stable for volumes lower than 7.58 Å$^3$/C. We interpret these results as suggesting that the bonding is $sp^3$-like for C-C bond lengths larger than 1.50 Å, while it is $sp^2$-like for smaller bond lengths. Within our calculations we have varied the C-C bond length and calculated the strain energy. Changing the bonding character does not affect the qualitative aspects of our results. We find the same dependence as the calculations carried out for the equilibrium C-C bond length in graphite, though we find a different prefactor for the $1/R^2$ dependence found earlier. The explanation for the changed prefactor comes naturally in an atomistic approach. As the C-C bond length is varied, the effective hopping matrix element between the Carbon atoms changes. The similar dependence found for the varied C-C bond lengths can also be understood. In any given system, the primary contribution to the strain energy comes from the sigma bonds which are bent in the process. The pi bonds are perpendicular to the sigma bonds remain grossly unaffected. Hence the completely atomistic approach will reflect the same trends as the continuum approach inspite of a change in the bonding character. While a change in the C-C bond length to larger values compared to the equilibrium value found for graphite should make the nanotube more stable than the sheet by virtue of the fact that the tube has more $sp^3$ bonding, a bond length variation from 1.2 Å to 1.8 Å does not find this to be the case.

**Methodology**

Carbon nanotubes are a single sheet of graphite that is rolled upon itself and connected at the edge. An Armchair nanotube is one classification of carbon nanotubes which corresponds to a configuration where there is no twist in addition to the rolling. We have carried out first principle electronic structure calculations for nxn armchair single-walled nanotubes where n was varied from 3 to 12. The nanotubes were taken to be open and extending to infinity in the z-direction. In the x- and y- directions, a large separation of 20 Å so as to minimize nanotube-nanotube interaction in a periodic system, and therefore simulate an isolated nanotube. For the tubes in which the C-C distance was larger (above 1.8 Å), we considered a separation of 26 Å. The plane wave pseudopotential implementation of density functional theory as implemented in the VASP code [5] was used for the calculations. The generalized gradient approximation [6] was used for the exchange functional. Ultrasoft pseudopotentials were used for Carbon in this study. A Monkhorst-Pack k-points set of 6x6x6 was used for the calculations of the total energies in each case. This was found to be sufficient as the prefactor $\alpha$, in the fitted expression $\alpha/R^2$ for the variation of the strain energy with radius, changes from 2.17 eV Å$^2$/atom to 2.02 eV Å$^2$/atom as the k-points is increased from 6x6x6 to 8x8x8. The strain energy was computed as the total energy difference between that of the carbon nanotubes of varying radii and the graphene sheet. The calculation was carried out for the equilibrium C-C bond length of 1.43 Å for graphite obtained by minimizing the total energy with respect to the volume as well as C-C bond lengths of 1.2 Å, 1.3 Å, 1.5 Å, and 1.8 Å. No geometry optimization was performed for the atoms making up the carbon nanotube.

**Results**

In Fig. 1 we have plotted the variation of the strain energy per atom calculated as a function of the nanotube radius. The C-C bond length in this case has been kept fixed at the value of 1.43 Å which corresponds to the equilibrium bond length for graphite found by us in our calculations. The strain energy is found to increase monotonically as the radius is decreased. This is expected as the distortion from the ideal graphite structure is more in the case of the smaller radii tubes. Continuum elasticity models, first pointed out by Tibbetts [4] predict a $1/R^2$ dependence of the strain energy. Fitting a $1/R^2$ dependence to the strain energy we extract a prefactor of 2.17 eV Å$^2$/atom for the dependence which is in agreement with the values obtained by other groups earlier. This value of the prefactor translates into a value of 2.14 TPa for the Young's modulus for the 10x10 nanotube when we take the thickness of the carbon nanotube to be 1.43 Å and the atomic volume to be 1.328 Å$^2$/atom. This is in reasonable agreement with the experimental values that are in the range of TPa for single wall nanotubes.

It has been shown earlier [7] that the strain energy dependence is well described considering only the electronic contribution within a tight-binding framework. Each C atom is coordinated to 3 other C atoms. To a first approximation, the changes in the C-C bondlengths is small and so the largest contribution to the strain energy comes the bending of the C-C-C angle from its ideal value from graphite. In the nanotube, for each C atom, it is found that 2 of the C-C-C angles deviate strongly from 120 degrees, while the third remains almost unchanged. Hence the strain energy can be related to a function of the form $A + B\cos^2\theta$.. These results have been plotted in Fig. 1. The prefactor B is related to the hopping matrix element between the carbon atoms in the nanotube, while the constant A is related to the hopping matrix element between the atoms in the graphite sheet. The dependence on the bending angle translates into a dependence on the radius.

The bonding at the bond length of 1.5 Å is primarily sp$^2$ type. We have varied the volume considering both graphite structure as well as the diamond structure for Carbon. The results for the total energy per atom are plotted as a function of the volume per atom in Fig. 3. For volumes lower than 7.58 Å$^3$/C we find that the diamond structure is more stable than graphite. The equilibrium bond length of the diamond structure was found to be 1.55 Å. We interpret this result as suggesting that the bonding will change over from sp$^2$ to sp$^3$ above this C-C bond length. We changed the C-C bond length from 1.2 Å to 1.8 Å. By doing this, we were trying to change the type of bonding. In all that cases we did not find any change in the dependence of the strain energy vs. R$^2$ or in the strain energy vs. cos θ curve.

Finally we examine how the values of A and B change as a function of the C-C bond-length. In Fig. 4 we have plotted the variation of the strain energy as a function of cos θ for four different values of the C-C bond length. Just as in Figs. 1 and 2, we are able to get a good fit within a continuum elasticity model as well as cos$^2$ θ . This was initially surprising as the bonding character was changing as the C-C bond length was being changed. The reason behind the success of the model is because it is primarily the contribution to the strain from the sigma bond that is changing as the bond length is being changed. The dependence of this bonding energy has the $\cos^2\theta$ dependence discussed earlier. Hence while the amount of sigma bond character might vary, the dependence remains the same. The prefactor of the $\cos^2\theta$ dependence, B, is found to decrease as the bond length is increased. This is indeed as expected as the hopping interaction strength between the Carbon atoms will decrease as the separation between them is increased.

**Conclusion**

To conclude, we have performed calculations for the strain energy involved in the formation of nanotubes from graphite sheets. The calculated result can be fit to both a continuum elasticity model as well as a model which captures the atomistic aspects. We show where atomistics enter the description of the strain energy.

FIGURES

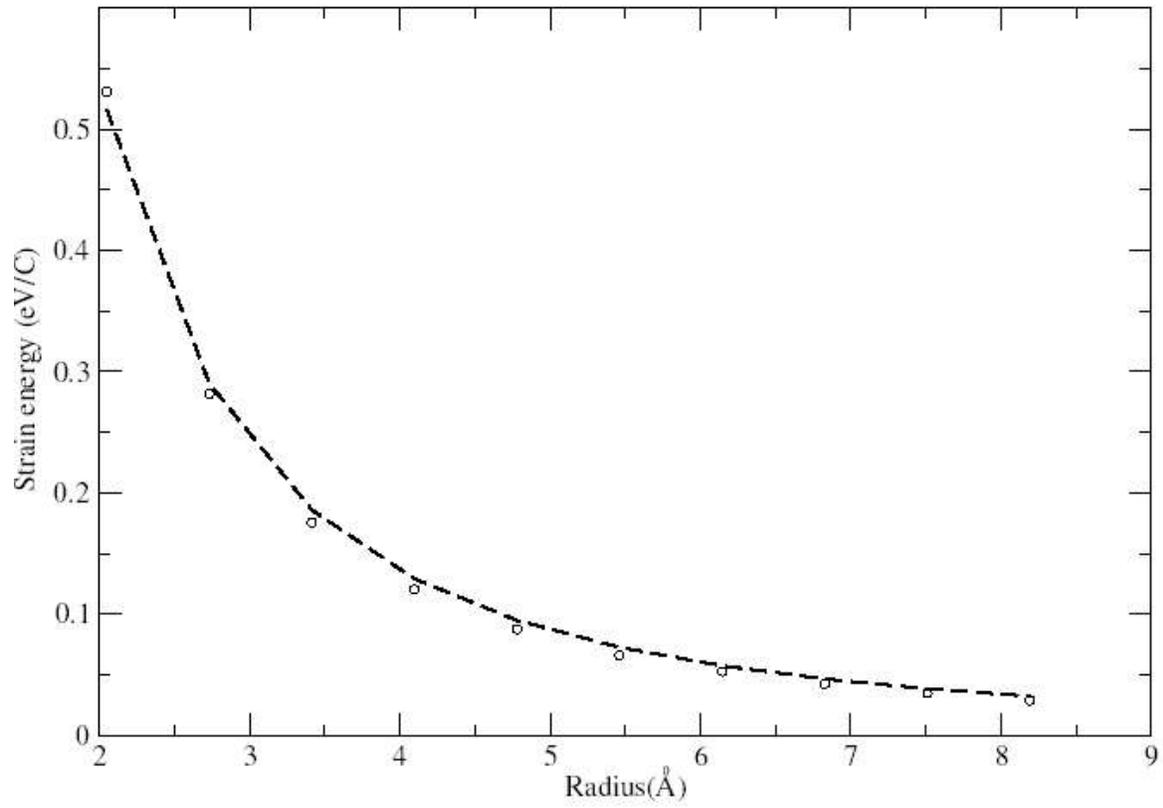

Fig. 1 : The variation of the calculated strain energy per atom (open circles) as a function of the nanotube radius (R). The dashed curve represents a fit to the relation $\alpha//R^2$

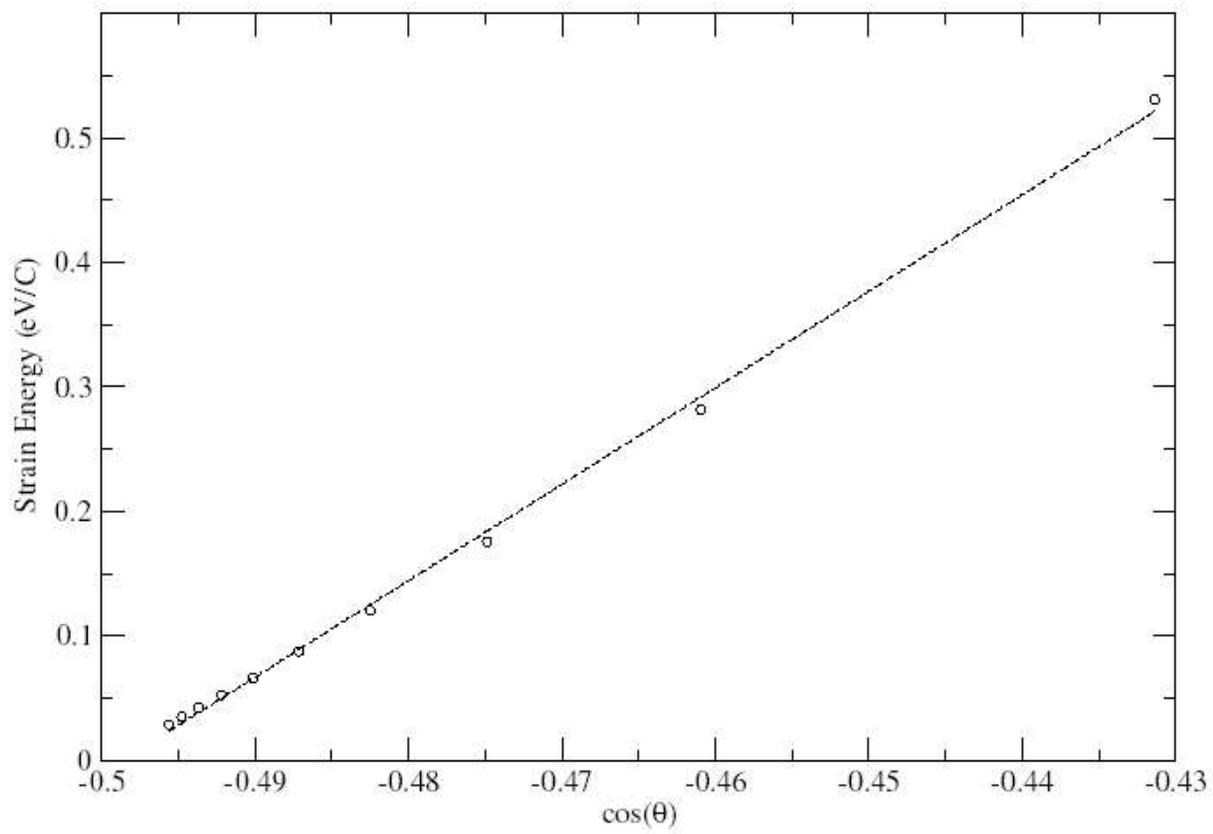

Fig. 2: Plot of the strain energy per atom (open circles) as a function of the C-C-C angle $\theta$ bent while forming the nanotubes. The dashed line represents the fit to the relation $A + B*\cos^2\theta$.

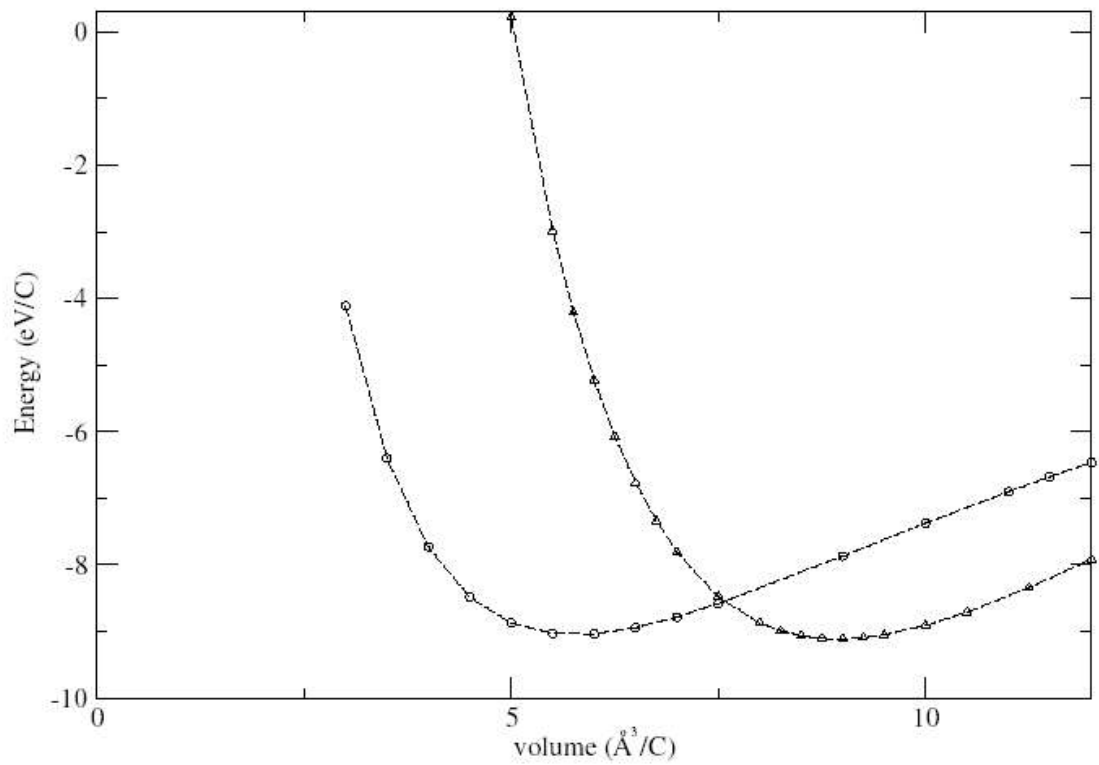

Fig. 3: The variation of the energy per C atom as a function of volume in graphite (open triangles) and diamond (open circles).

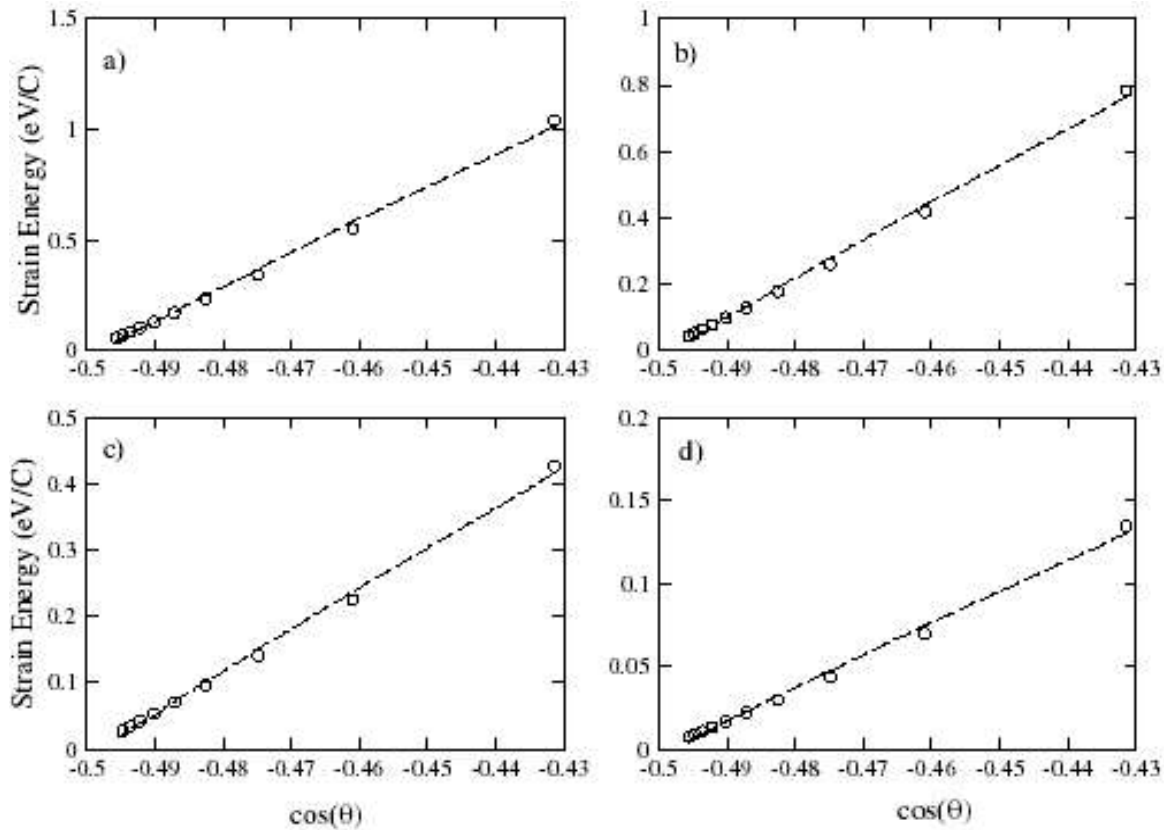

Fig. 4 : The variation of the strain energy per atom (squares) as a function of the C-C-C angle θ for C-C bond lengths of (a) 1.2 Å, (b) 1.3 Å, (c) 1.5 Å and (d) 1.8 Å. The dashed line corresponds to the fitted expression $A + B \cos^2\theta$. A is found to be 4.04, 3.05, 1.66 and 0.51 and B is found to be -16.296, -12.27, -6.71 and -2.09 for (a) – (d) respectively.